\documentclass[12pt,pdf,epsfig,color,png]{article}
\usepackage{amsmath}
\usepackage{amsfonts}
\usepackage{amsmath}
\usepackage{slashed}
\usepackage{amsxtra}
\usepackage{amstext}
\usepackage{amssymb}
\usepackage{color}
\usepackage{float}
\usepackage{graphicx}
\usepackage{subcaption}
\numberwithin{equation}{section}
\usepackage[dvips]{epsfig}
\usepackage{appendix}
\newcommand{\be}{\begin{equation}}
\newcommand{\ee}{\end{equation}}
\newcommand{\beq}{\begin{equation}}
\newcommand{\eeq}{\end{equation}}
\newcommand{\ba}{\begin{eqnarray}}
\newcommand{\ea}{\end{eqnarray}}
\newcommand{\bea}{\begin{eqnarray}}
\newcommand{\eea}{\end{eqnarray}}
\newcommand{\nn}{\nonumber}
\begin{document}
\baselineskip=15.5pt \pagestyle{plain} \setcounter{page}{1}
%
%--------+---------+---------+---------+---------+---------+---------+
%--------+---------+---------+---------+---------+---------+---------+
\begin{titlepage}

\vskip 0.8cm

\begin{center}
%
%
%
%--------+---------+---------+---------+---------+---------+---------+
%--------+---------+---------+---------+---------+---------+---------+
%\begin{titlepage}
%
%\vskip 0.8cm
%
%\begin{center}

{\Large \bf Proton helicity structure function $g_1^p$ from a holographic Pomeron}

\vskip 1.cm

{\large {{\bf Ignacio Borsa$^{a,}$}{\footnote{\tt iborsa@df.uba.ar}}, {\bf David Jorrin$^{b, c, }$}{\footnote{\tt jorrin@fisica.unlp.edu.ar}}, {\bf Rodolfo Sassot$^{a, }$}{\footnote{\tt sassot@df.uba.ar} {\bf and}}, \\ {\bf Martin
Schvellinger$^{b, c, }$}{\footnote{\tt martin@fisica.unlp.edu.ar}}}}

\vskip 1.cm

{\it $^a$ Universidad de Buenos Aires,  Facultad de Ciencias Exactas y Naturales, Departamento de F\'{\i}sica and IFIBA-CONICET, Ciudad Universitaria, Pabell\'on 1 (1428) Buenos Aires, Argentina.} \\

{\it $^b$ Instituto de F\'{\i}sica La Plata-UNLP-CONICET. 
Boulevard 113 e 63 y 64, (1900) La Plata, Buenos Aires, Argentina.} \\

{\it $^c$ Departamento de F\'{\i}sica, Facultad de Ciencias Exactas, Universidad Nacional de La Plata. Calle 49 y 115, C.C. 67, (1900) La Plata, Buenos Aires, Argentina.}

\vspace{1.cm}

{\bf Abstract}

\end{center}

\vspace{.5cm}

We present a detailed analysis of the polarized and the unpolarized deep inelastic scattering structure functions of the proton, $g_1^p$ and $F_2^p$ respectively, in the context of a holographic dual description based on type IIB superstring theory. We compare this description with experimental data and Quantum Chromodynamics estimates computed at leading, 
next-to-leading and next-to-next-to-leading order in perturbation. We confront the predictions of a holographic dual 
model and those of perturbative QCD for $g_1^p$ at the kinematics that will be probed by the forthcoming Electron-Ion Collider. We find that the extrapolation of $g_1^p$ to very small values the Bjorken variable computed with a Holographic Pomeron model based on actual data at higher momentum fractions is always positive and differs significantly with standard 
projections based on perturbative QCD. 

\noindent

\end{titlepage}

\newpage

{\small \tableofcontents}

\newpage

%\cite{Dokshitzer:1977sg}
%\cite{Gribov:1972ri}

%%%%%%%%%%%%%%%%%%%%%%%%%%%%%%%%%%%%%%%%%%%%%%%%%%%%%%%%%%%%%%%%

%---------------------------------------------------------------
%
\section{Introduction}
%
%---------------------------------------------------------------

Over the last fifty years our knowledge of the proton structure has deepened relentlessly. Deep inelastic scattering (DIS) experiments at SLAC \cite{Breidenbach:1969kd,Bloom:1969kc}
started showing hints of the scaling behavior that emerges from the asymptotic freedom of quarks already in the late sixties, triggering the concept of partons and the development of Quantum Chromodynamics (QCD) \cite{Altarelli:1981ax}, while HERA \cite{H1:2009pze} opened the current century testing with exquisite precision the departures from scaling and allowing 
to confront the data with the predictions of perturbative QCD, in particular in the Dokshitzer-Gribov-Lipatov-Altarelli-Parisi (DGLAP) framework \cite{Dokshitzer:1977sg,Gribov:1972ri,Altarelli:1977zs}. These estimates were laboriously 
developed and tested over decades in parallel with the experimental efforts and today the aim is to check the proton structure beyond the next-to-next-to-leading order (NNLO) 
accuracy \cite{Amoroso:2022eow}.

The remarkable success of the DGLAP approach reproducing the behavior of the data in a wide kinematic range certainly dazzled the community, perhaps veiling its provisional character as an approximation that at some point necessarily becomes inadequate. The forthcoming Electron-Ion Collider (EIC) \cite{AbdulKhalek:2021gbh} will dramatically extend our kinematic access and enhance the precision of the DIS measurements, thus driving us in that direction. Then, it is of the greatest interest to prepare ourselves for that contingency, for instance, producing well-motivated predictions that depart from the DGLAP scenario to complement impact studies and projections based mostly on the assumption of the validity of the DGLAP approximation.

In this respect, the Brower-Polchinski-Strassler-Tan (BPST) Pomeron approach provides a framework that reproduces with remarkable accuracy actual spin-independent structure function $F_2^p$ data with a deep and clear motivation together with a surprising economy of parameters. The BPST Pomeron was derived from type IIB superstring theory in curved spacetime, in the context of the gauge/string theory duality \cite{Brower:2006ea}. This Pomeron is a Regge trajectory of the graviton which carries the vacuum quantum numbers and is exchanged in the scattering process of four closed strings in the Regge limit. It allows to describe in a unified way both the perturbative Balitsky-Fadin-Kuraev-Lipatov 
(BFKL or hard) Pomeron (for negative values of the $t$-channel Mandelstam variable) and the soft Pomeron (for $t>0$). These situations occur in the $|t| \ll s$ limit, where $\sqrt{s}$ is the total energy of the system in the center-of-mass frame.

The BPST Pomeron approach was used to calculate the proton structure function $F_2^p$ and to fit HERA \cite{H1:2009pze} data with remarkable accuracy using only four free parameters \cite{Brower:2010wf}. Later, it was slightly modified to 
include also that of the H1-ZEUS \cite{H1:2015ubc}, BCDMS \cite{BCDMS:1989qop}, NMC \cite{NewMuon:1996fwh}, E665 \cite{E665:1996mob} and SLAC \cite{Whitlow:1991uw} collaborations within the ranges 0.1 GeV$^2 < Q^2 \leq 400$ GeV$^2$ and $2.43 \times 10^{-6} \leq x < 0.01$ \cite{Jorrin:2022lua}. The BPST Pomeron framework however extrapolates $F_2$ in a way that deviates from current DGLAP based fits to data, especially for very small and very large values of the photon virtuality $Q^2$, and for very small values of $x$. Of course, 
in the case of the estimates coming from DGLAP-based global fits to data, the low-$x$ extrapolation comes just from an assumption on the behavior of the parton distribution functions (PDFs) loosely motivated on the quark charge and momentum conservation and the simplest functional form required to fit the data at much larger $x$, whereas for the BPST Pomeron it is fixed by the model itself. In fact, in the formal derivation of the BPST Pomeron it is assumed that it holds for $x$ smaller than $1/\exp({\lambda_{\text{'t Hooft}}^{1/2}})$, where $\lambda_{\text{'t Hooft}} \gg 1$ is the 't Hooft coupling.

Furthermore, in the case of spin-dependent observables there is another construction also based on the gauge/string theory duality, that we call Holographic-$A$ Pomeron \cite{Kovensky:2018xxa}. This construction allows to parameterize the spin-dependent structure function $g_1^p$ in terms of three of the parameters fixed by $F_2^p$ data plus a single additional parameter which can be constrained by existing measurements of $g_1^p$ \cite{Kovensky:2018xxa,Jorrin:2022lua}. By Holographic-$A$ Pomeron in the following we specifically refer to the exchange of a Regge trajectory of a gauge field which in type IIB superstring theory is a linear combination of a gravi-photon and a fluctuation of the Ramond-Ramond four-form field $A_4$, firstly proposed and developed in \cite{Kovensky:2018xxa}. This object is different from the BPST Pomeron which exchanges the Reggeized graviton \cite{Brower:2006ea}, and from the Odderon which exchanges the Reggeized Kalb-Ramond field \cite{Brower:2008cy} of type IIB superstring theory. 

The Holographic-$A$ Pomeron reproduces $g_1^p$ data in the ranges $0.0036 \leq x < 0.01$ and $0.062$ GeV$^2< Q^2 < 2.41$ GeV$^2$ from SMC \cite{SpinMuon:1998eqa}, E143  \cite{E143:1998hbs}, COMPASS \cite{COMPASS:2010wkz,COMPASS:2015mhb,COMPASS:2017hef} and HERMES \cite{HERMES:2006jyl} 
collaborations, with great precision \cite{Jorrin:2022lua}. The extrapolation provides a prediction for $g_1^p$ at small $x$ in clear disagreement with DGLAP solutions that nevertheless
reproduce the data that is used to constrain the Holographic-$A$ Pomeron.

Taking into account realistic error estimates for the projected measurements of $g_1^p$ at the EIC \cite{Borsa:2020lsz} and the Holographic-$A$ Pomeron extrapolation to the small $x$ regime, 
it is then possible to assess if the EIC will be able to favor scenarios motivated by DGLAP dynamics, the Holographic-$A$ Pomeron or some other underlying physics. The history of the proton spin has always favored the unexpected \cite{Aidala:2012mv}.

In the next section we very briefly examine the path from string theory to DIS structure functions, introducing in a rather pedagogical manner what we mean by a dual holographic model and the role of the Pomeron. We defer a more detailed discussion for the interested reader to appendix A. Next, we revisit the phenomenology of the BPST Pomeron description of the unpolarized DIS structure function and show how it compares with the standard DGLAP picture. Finally, in the last section we examine the Holographic-$A$ Pomeron expectation for the spin-dependent structure function $g_1^p$ at the kinematics of the forthcoming Electron-Ion Collider, discuss how it compares with the projected errors and the most standard DGLAP projections.

%---------------------------------------------------------------
%
\section{String theory dual description of DIS at low $x$}
%
%---------------------------------------------------------------

The BPST Pomeron and the Holographic-$A$ Pomeron are both derived within the framework of the gauge/string theory duality.
This duality relates a non-Abelian gauge theory defined on a flat four-di\-men\-sio\-nal spacetime and superstring theory compactified on a certain ten-dimensional curved background \cite{Maldacena:1997re,Gubser:1998bc,Witten:1998qj}. The paradigmatic example is represented by the large $N_c$ limit of ${\cal {N}}=4$ supersymmetric Yang-Mills (SYM) theory with gauge group $SU(N_c)$ which, by the mechanisms of this duality, is related in a very specific way to type IIB supergravity on the AdS$_5 \times S^5$ background, which is an exact solution of the equations of motion of this supergravity. The radius of the five-dimensional sphere $S^5$ and the scale of the anti de Sitter (AdS) spacetime is a length given by $R=(4 \pi \lambda_{\text{'t Hooft}} \, \alpha'^2)^{1/4}$. The 't Hooft coupling is defined as $\lambda_{\text{'t Hooft}} \equiv g_{YM}^2 N_c$, being $g_{YM}$ the coupling constant of ${\cal {N}}=4$ SYM theory, and $\alpha'$ is the square of the fundamental string length. Recall that for the gauge theory one usually defines $\alpha_{\text {strong}} \equiv g_{YM}^2/4 \pi$.

The duality can be extended in many directions, for instance, one may consider the $1/N^2_c$ expansion of the gauge theory in terms of the genus expansion of the closed string world-sheet, where the genus counts the number of holes (or handles) that a two-dimensional closed surface contains. Thus in the large $N_c$ limit there are no holes, then the corresponding world-sheet is a two-dimensional sphere. Also, in the example presented above it is assumed the gauge theory to be strongly coupled, $1 \ll \lambda_{\text{'t Hooft}}$. This means that one must consider the low-energy limit of type IIB superstring theory, namely type IIB supergravity. Furthermore, one can go to finite coupling in the gauge field theory by considering an expansion in powers of $\alpha'$ (dual to the strong coupling expansion in powers of $\lambda_{\text{'t Hooft}}^{-1/2}$ on the gauge theory side), which implies that string theory states become dominant for the dynamics of the system. The duality bears a crucial property called the strong/weak coupling duality, which means that when the gauge theory is strongly coupled the associated dual string theory is weakly coupled, and reciprocally. Such property allows for a consistent description of a strongly coupled gauge theory in terms of a weakly coupled string theory dual model. This precisely permits to use it to investigate field theory processes for which non-perturbative dynamics becomes essential.

There is another key property inherent to the curved superstring theory background, and particularly when it includes the AdS spacetime. This comes from the so called warp factor multiplying the ``flat" four-dimensional piece of the metric, which induces a red-shift \cite{Polchinski:2001tt} as explained below. Let us consider the metric of the AdS$_5 \times S^5$ solution of type IIB superstring theory written in the following form
\begin{equation}
ds^2= \frac{r^2}{R^2} \,  \eta_{\mu\nu} dx^\mu dx^\nu + \frac{R^2}{r^2} dr^2 + R^2 d\Omega_5^2 \, , \label{metric1}
\end{equation}
with the radial coordinate $r$, which increases in the UV of the dual gauge theory. In the previous equation the last term ($R^2 d\Omega_5^2$) gives the piece of the metric corresponding to the five-sphere $S^5$, while the first two terms correspond to the $AdS_5$ space. It is usual to introduce an arbitrary IR cut-off at $r_0$ in the metric above, which induces color confinement in the dual gauge theory at the energy scale $\Lambda\equiv r_0/R^2$ \footnote{We work in natural units $c=\hslash=1$.}. In addition, the AdS$_5$ space has a boundary which is a four-dimensional Minkowski spacetime, whose indices are $\mu,\nu,\cdots=0,..., 3$. The conserved four-momentum $p^\mu_{4d}=-i \partial/\partial x_\mu$ is related to the ten-momentum $\tilde{P}^\mu_{10d}$ in local inertial coordinates at certain point $r$ of the AdS$_5$ space as follows
\begin{equation}
p^\mu_{4d} = \frac{r}{R} \, \tilde{P}^\mu_{10d} \, .
\end{equation}
Therefore, a string theory scattering process localized at the position $r$ within the AdS$_5 \times S^5$ spacetime corresponds to a particle scattering process with four-momentum $p^\mu_{4d}$ from the gauge theory perspective. Thus, as $r$ decreases in the bulk of the AdS space it corresponds to a process in the IR of the gauge theory. These ideas were applied to hard scattering in \cite{Polchinski:2001tt} and to deep inelastic scattering of glueballs and fermions in \cite{Polchinski:2002jw}. In particular, for low values of the Bjorken variable Brower, Polchinski, Strassler and Tan \cite{Brower:2006ea} developed the BPST Pomeron, which is the gauge/string theory dual object which unifies the (soft) Regge and the (hard) BFKL Pomerons. The BPST Pomeron describes very well the structure function $F_2^p$ of the proton at low $x$ \cite{Brower:2010wf}. On the other hand, there is the Holographic-$A$ Pomeron \cite{Kovensky:2018xxa} which describes very well the existing experimental data of the proton helicity structure function $g_1^p$ at low $x$ \cite{Jorrin:2022lua}. \\

Before introducing the BPST and the Holographic-$A$ Pomerons, we will very briefly remind what are the soft and hard Pomerons. The idea is to make connexions between the previous $S$ matrix and gauge theory approaches and the more recent gauge/string theory duality perspective. A more detailed description is presented in appendix A.

Almost a decade before the introduction of the QCD Lagrangian, the extraordinarily challenging problem of describing strong interactions was investigated using the $S$-matrix framework.
This led to the so-called Regge theory, which was used to study the cross-sections of hadron-hadron and photon-hadron scattering processes at high energy \cite{Devenish:2004pb},
borrowing concepts from potential scattering in quantum mechanics but enforcing 
%There are three very important postulates about the $S$ matrix, namely:
 Lorentz invariance, unitarity and analyticity \cite{Forshaw:1997dc,Donnachie:2002en}.
Let us consider a two-to-two particles scattering process, with incoming particles $i_1$ and $i_2$ and the outgoing ones $f_3$ and $f_4$. The incoming four-momenta are $p_1^\mu$ and $p_2^\mu$ and the outgoing four-momenta are $p_3^\mu$ and $p_4^\mu$, while their masses are $m_j$ ($j=1, \cdots, 4 $), respectively. This process can be described in terms of the Mandelstam variables:
\begin{equation}
s = (p_1+p_2)^2 \, , \,\,\,\,\,\,\,
t = (p_1-p_3)^2 \, , \,\,\,\,\,\,\,
u = (p_1-p_4)^2 \, . \label{stu}
\end{equation}
being $t$ the square of the four-momentum exchanged between particles $i_1$ and $f_3$, and there is also the kinematic relation $s+t+u=\sum_{j=1}^4 m_j^2$. Therefore, the transition amplitude for the process $i_1+i_2 \rightarrow f_3 + f_4$ is a function of only two independent Mandelstam variables, ${\cal {A}}(s, t)$. The study of this scattering amplitude suggests that there is the exchange of an object carrying angular momentum which is a function of the Mandelstam variable $t$ (say $j=\alpha(t)$), called Reggeon, which is not a single particle. Therefore, this scattering amplitude can be interpreted as the superposition of amplitudes corresponding to the exchanges of all possible particles in the $t$-channel, which leads to a Regge trajectory. Moreover, for positive values of the Mandelstam variable $t$, experimental data show that the scattering amplitude must be dominated by the exchange of a Reggeon with zero isospin, which has to be even under charge conjugation. This particular Reggeon is called the soft Pomeron. 
The connexion with the symmetric structure functions $F_1$ and $F_2$ comes from the fact that DIS cross-section can be written in terms of the $\gamma^* + p$ scattering process by using the optical theorem, where $\gamma^*$ represents a virtual photon, with squared four-momentum $q^2=-Q^2$. At low $x$ the behavior of the total cross-section of a virtual photon-proton scattering is dominated by the exchange of a Pomeron, leading to $F_2(x, Q^2) \propto x^{-0.08}$ as the Bjorken variable goes to zero.

There is another Pomeron, called hard or BFKL Pomeron, which has been derived from QCD in perturbation theory. The lowest order Feynman diagram from QCD which perturbatively can simulate a Pomeron exchange like this is given by a two-gluon exchange. This Pomeron is derived from the BFKL equation \cite{Fadin:1975cb,Kuraev:1976ge,Kuraev:1977fs,Balitsky:1978ic}. The problem is still how to calculate the proton impact factor, for which one may try different models. On the other hand, there is an issue due to that in QCD the next-order correction to the BFKL Pomeron is large and has an opposite sign with respect to the single BFKL Pomeron itself \cite{Fadin:1998py,Camici:1997ij}.

As described in the introduction, for certain hadron scattering processes at high energy ($s \gg |t| \gg \Lambda^2_{QCD}$, where $\Lambda_{QCD}$ is the IR scale of QCD) and small scattering angle the Regge theory suggests the exchange of a soft Pomeron (Reggeon) for positive $t$ values, and a single BFKL-Pomeron exchange at leading order in $\alpha_{\text{strong}} \log s$ at weakly coupled QCD for $t \leq 0$. The soft Pomeron is understood as an exchange of a single glueball, which in the string theory dual language corresponds to a closed string. On the other hand, the BFKL framework entails the exchange of a color-singlet object composed by Reggeized gluons, which is the BFKL Pomeron. Many aspects of QCD simplify when one considers the large $N_c$ limit, where $N_c$ is the rank of the gauge group $SU(N_c)$.  In the present context the large $N_c$ limit implies that the dominant contribution to the scattering amplitude comes from a single Pomeron exchange. From the type IIB superstring theory perspective the dual exchanged object is a Reggeized graviton, leading to the BPST Pomeron \cite{Brower:2006ea}. The BPST Pomeron has a very important property, namely: at strong coupling of the gauge theory it unifies the soft and hard Pomerons, something which technically is not possible in QCD. In this context, Brower, Djuric, Sarcevic and Tan \cite{Brower:2010wf} obtained the structure function $F_2$ derived from the BPST Pomeron. This function has four free parameters, namely: $g_0^{2}$, $\rho$, $z_0$ and $Q'$ which are obtained by fitting it to experimental data as shown later, and it is given by
\begin{equation}
F^{\text{BPST}_{\text{HW}}}_2(x, Q^2) = \frac{g_0^2 \ \rho^{3/2} \ Q}{32 \ \pi^{5/2} \ \tau_b^{1/2} \ Q'} \ e^{(1-\rho) \tau_b} \left(e^{-\frac{\log^2{(Q/Q')}}{\rho \tau_b}}+ {\cal{F}}(x, Q, Q') \ e^{-\frac{\log^2{(Q Q' z^2_0)}}{\rho \tau_b}}\right) \, .
\label{FBPSTHW}
\end{equation}
The definition of the function ${\cal{F}}(x, Q, Q')$ as well as the physical meaning of the four parameters entering the above equation are given in appendix A.

Now, we turn the attention to the $g_1$ helicity function.
Although, QCD and ${\cal {N}}=4$ SYM are different theories, one should keep in mind the fact that within the parametric regimes of the momentum transfer and the Bjorken variable that we investigate here, the main contribution in both theories to the DIS process comes from the gluonic sector, which is similar in both theories. In this sense the behavior of the holographic Pomerons, both the BPST and the Holographic-$A$ Pomerons, is universal. In both situations the model dependence is related to the IR deformation and the hadron impact factor.

In the work \cite{Kovensky:2018xxa} it has been obtained the helicity structure function $g_1$ given by the following expression
\begin{equation}
g^{{\text{A}}_4 {\text{Pomeron}}_{\text{HW}}}_1(x, Q^2)  = \frac{C \rho^{-1/2} \ e^{(1-\frac{\rho}{4}) \tau_b}}{\tau_b^{1/2}} \left(e^{-\frac{\log^2{(Q/Q')}}{\rho \tau_b}}+ {\cal{F}}(x, Q, Q') \ e^{-\frac{\log^2{(Q Q' z^2_0)}}{\rho \tau_b}}\right)  \, .
    \label{g1APomeron}
\end{equation}
Notice that the parameters $\rho$, $Q'$ and $z_0$ should be fixed by the fitting of $F^{\text{BPST}_{\text{HW}}}_2(x, Q^2)$ to experimental data, since the physical meaning of them is the same in both structure functions. Then, there is only one free parameter to fit to $g_1^p$ experimental data, the overall constant $C$. Details are explained in appendix A.

%---------------------------------------------------------------
%
\section{$F_2^P$ structure function}
%
%---------------------------------------------------------------

Before discussing the polarized structure function, in this section we revisit the unpolarized structure function $F_2^p$ to remind how good is the agreement of the BPST Pomeron picture with data and to show how it compares to DGLAP-based estimates. As it was mentioned above and discussed in detail in \cite{Jorrin:2022lua}, three of the four parameters that determine the behavior $g_1^p$ in the Holographic-$A$ Pomeron approach are associated with the BPST Pomeron model for $F_2^p$, so it is also a cornerstone for the spin-dependent results.

\begin{figure}[H]
%\centering
%\includegraphics[scale=0.8]{Fig_compl_x_sx.eps}
\hspace{-0.3cm}
\includegraphics[width=3.2in]{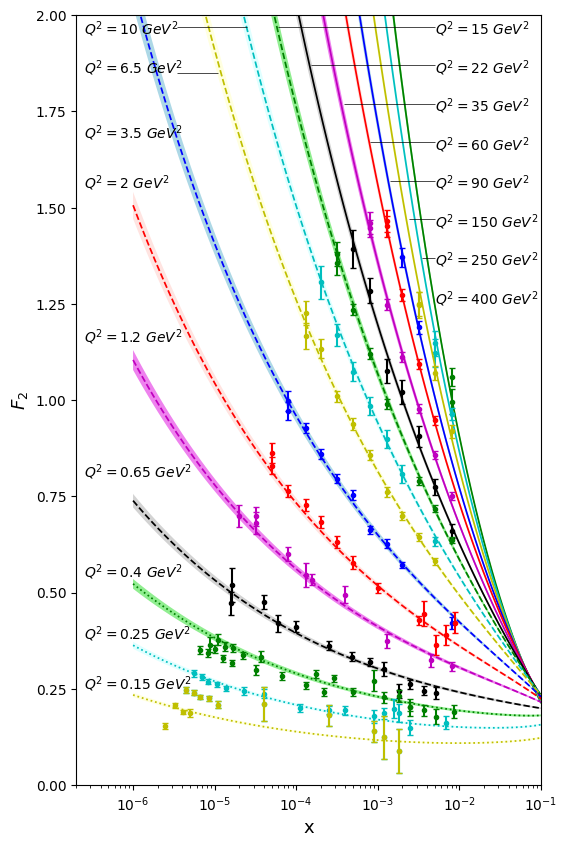}
\hspace{0.4cm}
\includegraphics[width=3.09in]{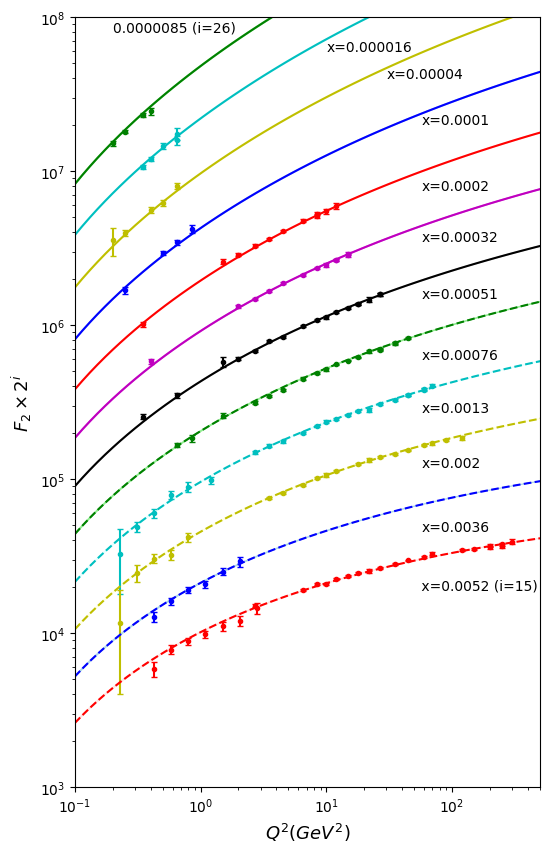}
\caption[ ]
{\small The proton $F_2^p$ structure function using a single BPST Pomeron exchange against data from H1-ZEUS, BCDMS, NMC, E665 and SLAC collaborations within the ranges 0.1 GeV$^2 < Q^2 \leq 400$ GeV$^2$ and $2.43 \times 10^{-6} \leq x < 0.01$.  The number of  data points depicted has been limited for a better visualization. Error bands are included in both figures. Due to the logarithmic vertical scale in the right hand side plot, though the error bands are present, they are very narrow and cannot be distinguished from their central values.} 
\label{Fig1}
\end{figure}

In Figure 1 we show the unpolarized structure function $F_2^p$ both as a function of the Bjorken variable $x$ (left hand side plot) and the photon virtuality $Q^2$ (right hand side plot) respectively. The curves result from fitting the four BPST Pomeron parameters to 280 data points from DIS experiments with a resulting $\chi^2_{\text{d.o.f.}}$ of 1.086, that reflects the  quite remarkable agreement. The values of the parameters are:
\begin{eqnarray}
&& \rho=0.7729 \pm 0.0014, \,\,\,\,\,\,\, \,\,\,\,\,\,\,\,\,\,\,\,\,\,\,\,\,\,\,  g_0^2= 103.73 \pm 0.757, \nonumber \\
&& z_0=4.894 \pm 0.061 \, {\text {GeV}}^{-1},  \,\,\,\,\,\,\,\,\,\,\, Q'=0.4715 \pm 0.0093 \, {\text {GeV}}.  \label{F2parameters}
\end{eqnarray}
In this case it has been used a sieving method which excludes "outliers" with a $\Delta \chi^2_{\text {max}} = 4$ \cite{Jorrin:2022lua}.
Although the fit covers in principle $2.43 \times 10^{-6} \leq x < 0.01$ and 0.1 GeV$^2< Q^2 \leq 400$ GeV$^2$, it is clear from the plot that, as usual with DIS data, the data at lower $x$ correspond to extremely low $Q^2$ data, while higher $Q^2$ data points correspond to a rather limited range in the high values of $x$. The left hand side plot in Figure 1 emphasizes how well the BPST Pomeron picture reproduces the low-$Q^2$ behavior of the structure function, even for values well below 1 GeV$^2$, while the scale dependence at lower $x$ is not constrained by data at that region. This raises the question on how well the model behaves at low $x$ but higher $Q^2$ -the upper left corner of Figure 1- question that will certainly be answered by EIC. In the meantime, it is instructive to compare these BPST Pomeron expectations with the estimates for $F_2^p$ derived from parton distribution functions obtained in global QCD fits to data based on DGLAP dynamics. 

More specifically, in the DGLAP approximation the structure function $F_2^p$ is written as a convolution between coefficient functions $C_i^{(n)}(x,Q^2)$ that can be computed to a given order $n$ in perturbation theory for each parton type $i$, and non-perturbative but universal PDFs $f_i^{(n)}(x,Q^2)$ for the 
different parton types $i$, that are extracted from experiment within a perturbative approximation $n$ \cite{Altarelli:1977zs}. Schematically, 
\begin{equation}
F_2^p(x,Q^2)= \sum_i \int_x^1 \frac{dy}{y}\, C_i^{(n)} \left( \frac{x}{y}, Q^2 \right) \, f_i^{(n)}(y,Q^2) \, . 
\end{equation}
Even though the $x$ dependence of the PDFs cannot be computed from first principles in perturbation theory their $Q^2$ dependence is driven by the DGLAP equations, whose kernels $P_{ij}^{(n)}(x)$ can also be computed at a given order in perturbation   
\begin{equation}
\frac{d\,f_i^{(n)}(x,Q^2)}{d\, \log Q^2}=\frac{\alpha_s(Q^2)}{2\pi} \sum_j \int_x^1 \frac{dy}{y}\, P_{ij}^{(n)}\left(\frac{x}{y}\right)\,f_j^{(n)}(y,Q^2) \, .
\end{equation}

PDFs global analyses are not only based on DIS data, but are constrained and refined with information obtained from proton-proton collisions cross sections for a variety of final states 
\cite{Harland-Lang:2014zoa,NNPDF:2021njg}. Since PDFs are in 
turn an essential ingredient to analyze and interpret the results from collider data in the validation of the Standard Model and the searches of physics beyond it, a significant effort has been put in the last three decades to improve and refine them. Any physical observable, and in particular the DIS structure functions, can be computed from PDFs assuming factorization and universality in the leading twist and 
the leading logarithmic approximation (LO), as well as in the two following orders: next-to-leading logarithmic order (NLO) and next-to-next-to-leading  order (NNLO) in perturbation. These have been checked to be a very good approximations for inclusive DIS cross sections at intermediate values of $x$ and for increasing photon virtualities, starting at a few GeV$^2$. Below that limit, these approximations are expected to breakdown, and for this reason PDF global analyses are unable to exploit or predict DIS data there. Roughly speaking, the data points below the dashed purple line in the left hand side of Figure 1, are beyond the reach of the DGLAP approximations, but are nicely reproduced by the BPST Pomeron approach. Conversely, the DGLAP approach is expected to evolve faithfully to higher scales PDFs that are known at a lower one, precisely where the BPST Pomeron estimate becomes uncertain. A similar discussion is inferred from the behavior of $F_2^p$ as a function of $Q^2$ for different values of $x$ shown in the right hand side plot of Figure 1.

In Figure 2 we show the ratios between the LO, NLO and NNLO DGLAP-based estimates for $F_2^p$ and the BPST Pomeron parameterization mentioned above \cite{Jorrin:2022lua} and used in Figure 1. On the left hand side the plot shows the ratios as function of $x$ for fixed values of $Q^2$, and as function of $Q^2$ for fixed $x$ on the right. The DGLAP structure functions are computed using the NNPDF4.0 set of spin-independent PDFs from reference 
\cite{NNPDF:2021njg}. Entirely similar results are obtained with other modern PDFs sets provided $Q^2>3$ GeV$^2$. Beyond the LO approximation, modern sets of PDFs typically agree to 
a percent level in most of the kinematic range covered by the plots \cite{Harland-Lang:2014zoa}.

The bands around the curves in Figure 2 represent the estimated errors in the structure functions propagated from those of the PDFs for the DGLAP estimates, relative to the BPST estimate,  whereas the central (almost invisible) grey band is the relative error of the BPST Pomeron estimate propagated from that of their parameters. The bands reflect in part the uncertainty of the data used to extract the PDFs in the different kinematics, and also the error introduced by the different perturbative
approximations used in the PDFs extraction. Notice that the different perturbative approximations assume different $x$ and $Q^2$ dependence through the coefficient functions and evolution equations, therefore the lowest-order approximations presumably will be less able to accommodate data from different observables at different scales and momentum fractions, thus resulting in larger uncertainties as shown in the plots.

Starting with the bottom of the plot in the right hand side of Figure 2, we see that the BPST Pomeron and the three DGLAP estimates agree nicely for $x=0.01$  for $Q^2 > 10$ GeV$^2$ as one would expect, since we are well within the perturbative regime and the PDFs are strongly constrained by data. Of course, the three perturbative estimates assume slightly different scale dependence which become apparent at lower $Q^2$. The NNLO estimate (red line) is the one that remains closer to the BPST Pomeron for decreasing values of the scale, even down to $Q^2  \sim 2$ GeV$^2$. On the other hand, NLO estimate (green line) shows slightly poorer agreement, and the LO in light blue shows the largest difference. In the low $Q^2$ region is where the BPST Pomeron can be considered the most faithful estimate, since as we have already seen in Figure 1, it reproduces data down to a fraction of a GeV. On the other hand, PDFs are poorly constrained below a couple of GeV$^2$, and in fact at these low $Q^2$ values one can find large discrepancies between the results of different groups even in the NNLO approximation.

\begin{figure}[H]
%\centering
%\includegraphics[scale=0.8]{Fig_compl_x_sx.eps}
%\hspace{-2.9cm}
\hspace{-0.5cm}
%\vspace{1cm}
%\includegraphics[width=8.70in]{ratiosQ2.pdf}
\includegraphics[width=7.0in]{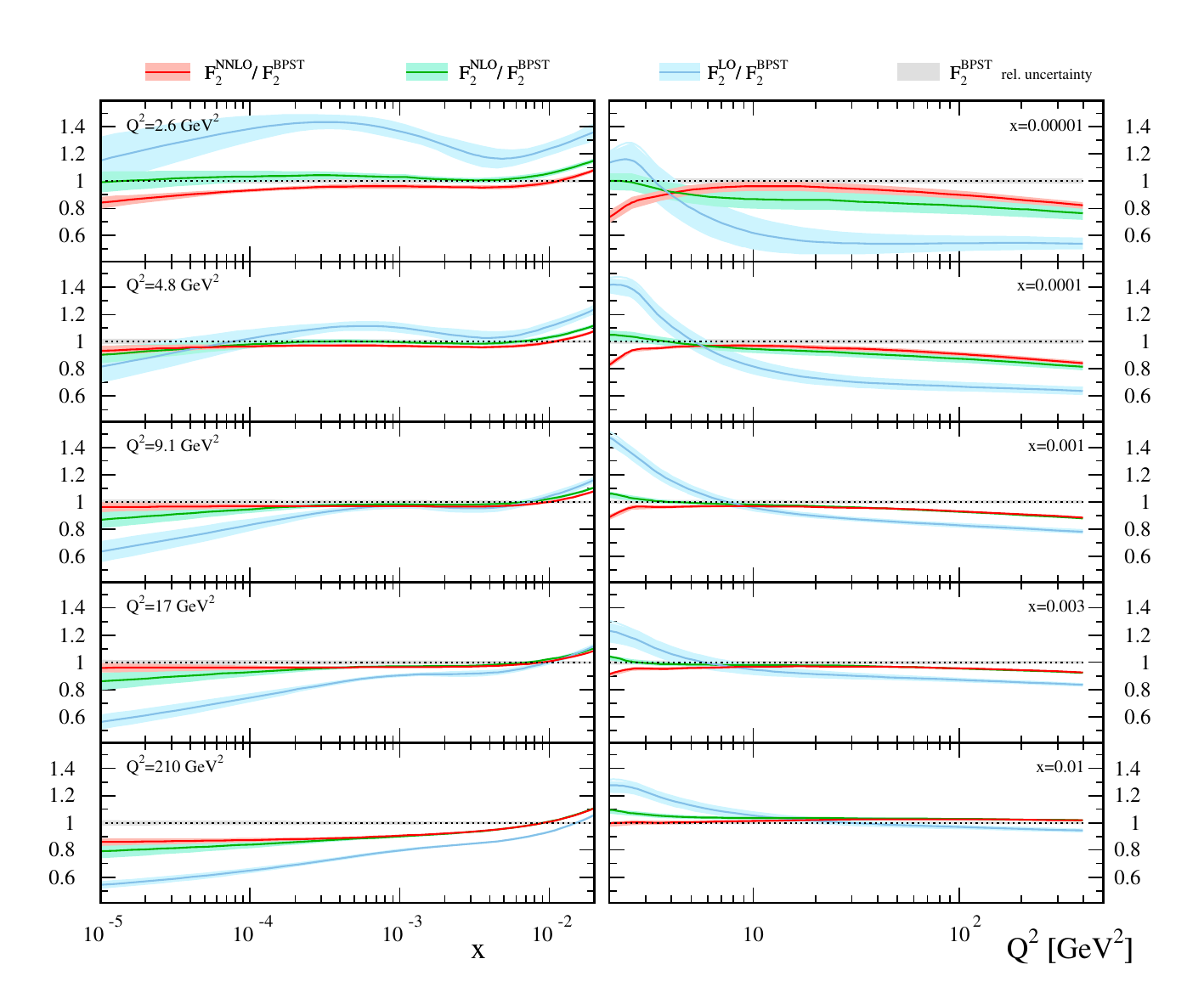}
\caption[ ]
{\small The ratios between the LO, NLO and NNLO DGLAP-based estimates for $F_2^p$ and the BPST Pomeron parametrization.} 
\label{Fig2}
\end{figure}

Going up in the right hand side plot of Figure 2, we reduce the value of the momentum fraction $x$, and we see that in addition to an increasing discrepancy between the three perturbative estimates at low $Q^2$, the LO differs also at higher values of $Q^2$ with the other estimates. Most likely this happens because the LO PDFs try to compensate the deficiencies in the $x$ and $Q^2$ dependencies of the coefficients mimicking the data with the strongest constraining power that typically correspond to larger $x$, at the expense of the less precise data at smaller $x$. The NLO and NNLO approximations have much more success connecting lower and higher $x$ data. It is interesting to notice that the perturbative convergence, roughly represented by the distance between the curves, is rather good beyond the NLO but decreases with decreasing $x$ as well as decreasing $Q^2$.

A crucial feature for our discussion in the next section on the helicity-dependent structure function is the remarkable agreement between the NNLO approximation and the BPST Pomeron estimate at $Q^2\sim 10$ GeV$^2$ and $x \sim 10^{-5}$, as shown in the top of the right hand side plot of Figure 2. From the point of view of the BPST Pomeron approach, the estimate in this kinematic regime is essentially and extrapolation, since there is no data on $F_2^p$ validating the model, as shown in Figure 1. Nevertheless, the BPST Pomeron estimate agrees remarkably well with the best perturbative estimate, even up to values of $Q^2\sim 20$ GeV$^2$. The importance of this feature lays in the fact that we will use this framework, and specifically three parameters of the BPST Pomeron $F_2^p$ in order to fix three of the four parameters of the Holographic-$A$ Pomeron, to make predictions for EIC for $g_1^p$ in this kinematics. For larger values of $Q^2$ the agreement clearly deteriorates; there, one expects the BPST Pomeron approach to be even less constrained while DGLAP is in good standing.

The plot on the left hand side of Figure 2 shows the same as that on right but now as a function of $x$, and emphasizing complementary aspects. The best overall agreement here takes place at an intermediate value of $Q^2 \sim 10$ GeV$^2$ between the NNLO and the BPST Pomeron estimate for almost all the range in $x$. At the largest values of $x$  ($x \sim 0.01$) where the PDFs are best constrained, but the BPST Pomeron is not expected to be a good approximation, predictably the agreement deteriorates.  Towards smaller $x$ the lower order approximations become increasingly inaccurate. Moving up in the plot towards lower $Q^2$, the perturbative predictions loose consistency between themselves, while in the opposite direction at increasing values of $Q^2$, the disagreement remains at small $x$. At the highest value of $Q^2$ in the bottom of the plot there is a sizable disagreement between the BPST Pomeron estimate and the NNLO in almost all the range of values of $x$.

We have explored the alternative of feeding the BPST Pomeron parameter determination with pseudodata on $F_2^p$ generated from the DGLAP projections to complement the DIS actual data set beyond the kinematical range accessible at present. However, the quality of the fits deteriorates significantly as more pseudodata at higher $Q^2$ is incorporated.

%\newpage

%---------------------------------------------------------------
%
\section{$g_1^p$ helicity-dependent structure function}
%
%---------------------------------------------------------------

In this section we focus on the helicity-dependent structure function of the proton $g_1^p$ whose measurements have received a great deal of attention since the EMC collaboration at CERN reported at the end of the eighties results consistent with a picture where very little of the proton spin came from the spin of the quarks, in contradiction with the naive quarks model \cite{EuropeanMuon:1989yki}. The EMC results were later confirmed by other DIS experiments, and more recently by measurements of final state jets and  hadrons in polarized proton-proton collisions at the Relativistic Heavy Ion Collider (RHIC) \cite{Aidala:2012mv}. The latter specifically showed that indeed a sizable contribution to the proton spin came from the polarization of gluons \cite{deFlorian:2014yva,Nocera:2014gqa}. The gluon polarization contributes to $g_1^p$ structure function albeit through terms suppressed by a power $\alpha_{\text{strong}}$ relative to those of the quark contributions, and also indirectly through the scale dependence of the quark contributions, which are coupled to the gluons by the spin-dependent DGLAP equations.

As in the unpolarized case, the helicity-dependent structure function $g_1^p$ can be written as a convolution between the appropriate perturbative spin dependent coefficient functions $\Delta C_i^{(n)}(x,Q^2)$ and spin-dependent or helicity PDFs $\Delta f_i^{(n)}(x,Q^2)$ \cite{Altarelli:1977zs}
\begin{equation}
g_1^p(x,Q^2)= \sum_i \int_x^1 \frac{dy}{y}\, \Delta C_i^{(n)}\left(\frac{x}{y},Q^2 \right) \, \Delta f_i^{(n)}(y,Q^2) \, , 
\end{equation}
where the latter are defined as the difference between the PDFs of partons with spin orientation parallel and antiparallel to that of the proton, i.e.
\begin{equation}
\Delta f_i(x,Q^2) \equiv f_i^{\uparrow}(x,Q^2)-f_i^{\downarrow}(x,Q^2) \, ,
\end{equation}
and that also obey evolution equations
\begin{equation}
\frac{d\,\Delta f_i^{(n)}(x,Q^2)}{d\, \log Q^2}=\frac{\alpha_{\text{strong}}(Q^2)}{2\pi} \sum_j \int_x^1 \frac{dy}{y}\, \Delta P_{ij}^{(n)}\left(\frac{x}{y} \right)\,\Delta f_j^{(n)}(y,Q^2) \, .
\end{equation}

Unlike the data on the unpolarized structure function $F_2^p$, the data on $g_1^p$ are much less precise and comparatively scarce, specially at low momentum fractions. Helicity-dependent PDFs obtained from DGLAP global analyses in turn inherit these shortcomings, redoubled by the fact there are no charge or momentum conservation for helicity distributions as in the unpolarized case, and that for the moment they only reach NLO precision. Therefore, the helicity distributions below $x\sim 10^{-3}$ are essentially extrapolations and their uncertainties, 
as well as, those for the spin-dependent structure functions in that regime are almost unbound. 

Again, precisely where the estimates for the structure function coming from DGLAP global analyses are more uncertain is where the string theory dual description is best constrained. Recall that the Holographic-$A$ Pomeron fits 56 data points on $g_1^p$ in the range $0.0036 \leq x \leq 0.009$ and $0.062$ GeV$^2 < Q^2 < 2.41$ GeV$^2$, adding just one free parameter to those already constrained by $F_2^p$ (see equations (\ref{FBPSTHW}) and (\ref{F2parameters})) with remarkable accuracy ($\chi^2_{d.o.f.}=1.14$). The referred parameter is the overall constant in the expression (\ref{g1APomeron})
\begin{equation}
C = 0.145 \pm 0.0015 \, . \label{Cconstant}
\end{equation}
Notice that we have not used any sieving for the experimental data of $g_1^p$, thus it includes all available data for the helicity structure function of the proton.
It seems natural to extrapolate this result to lower values of $x$ and moderate values of $Q^2$ for which we showed in the previous section that the BPST Pomeron picture reproduces the unpolarized structure function data in a very good approximation. One should emphasize that all the parameters for both the BPST Pomeron and for the Holographic-$A$ Pomeron are, in principle, independent on the Bjorken variable and the photon virtuality.

\begin{figure}[H]
\centering
\includegraphics[scale=0.8]{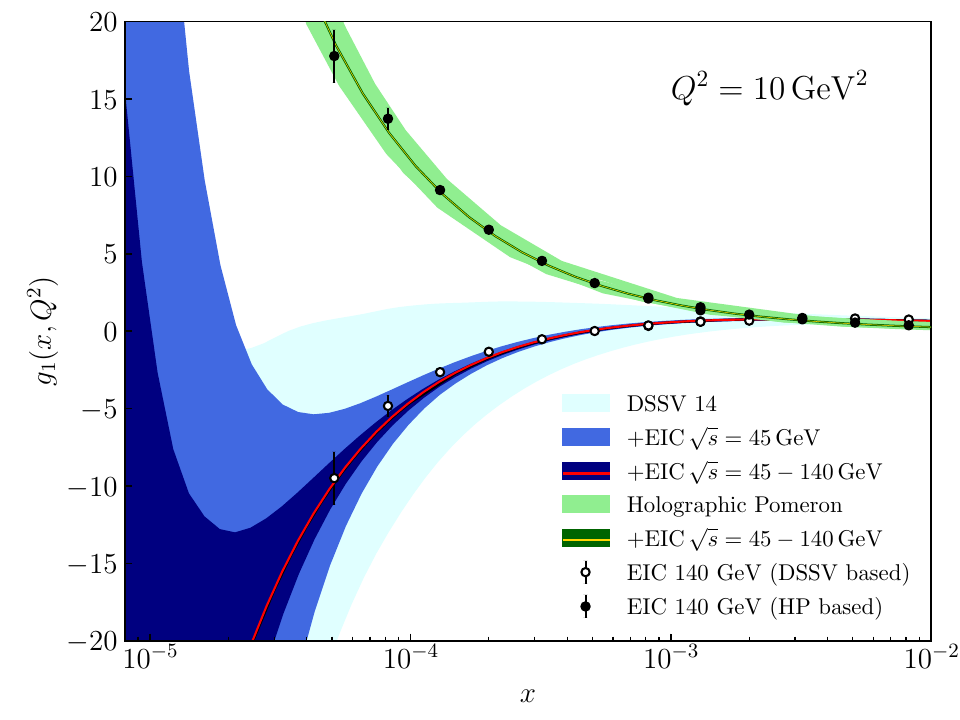}
\caption[ ]
{\small $g_1$ structure function using a single Holographic-$A$ Pomeron exchange to fit experimental data within the range  $0.0036  \leq x \leq 0.009$ at  $Q^2=10$ GeV$^2$ against the one obtained in DSSV14 DGLAP NLO analysis.} 
        \label{Fig3}
    \end{figure}
Interestingly, the extrapolation to low $x$ of the $g_1^p$ estimate coming from the Holographic-$A$ Pomeron differs
dramatically with those coming from most DGLAP helicity fits, like DSSV14 \cite{deFlorian:2014yva} shown in Figure 3. 
While the DSSV14 low-$x$ extrapolation for $g_1^p$ (in red) is increasingly negative, the Holographic-$A$ Pomeron result (green) goes in the opposite direction. The light-blue band represents the estimated uncertainty for the DSSV result, derived from the errors of the DSSV14 NLO DGLAP helicity PDFs \cite{DeFlorian:2019xxt}, while the light-green one is the one propagated from the Holographic-$A$ Pomeron and using three of the BPST Pomeron parameters. It is worthwhile noticing that in both approaches, the data on $g_1^p$ analyzed start at $x \geq 0.0036$ and consequently the uncertainty bands start growing there very fast towards smaller $x$.  In the case of the DGLAP approach, in principle $g_1^p$ could become positive at smaller values of $x$, however global analyses using simple functional 
forms for the helicity distributions prefer the negative solution. For the Holographic-$A$ Pomeron is it not possible to produce a negative $g_1^p$ compatible with the parameters $z_0$, $\rho$ and $Q'$ obtained from fitting $F_2^p$ to experimental data. This comes from the fact that the Holographic-$A$ Pomeron kernel has the same structure and signature as the BPST Pomeron as it can be seen by comparing equations (\ref{FBPSTHW}) and (\ref{g1APomeron}).

In reference \cite{Borsa:2020lsz} it has been argued that in a  scenario where the contributions from the gluon polarization to  $g_1^p$ dominate over those of quarks, a negative $\partial g_1^p/ \partial \ln Q^2$ corresponds to a positive gluon polarization that tends to compensate the smallness of the quark contribution to the spin of the proton. Conversely, a positive $\partial g_1^p/ \partial \ln Q^2$ represents negative gluon polarization that aggravates the deficit in the spin budget and favors more significant contributions from the angular momentum, for example. In this respect, 
the Holographic-$A$ Pomeron solution clearly favors the latter as it can be seen in Figure 10 of reference \cite{Jorrin:2022lua}. Interestingly, in refeference 
\cite{Adamiak:2023yhz}  it has been shown that within the Kovchegov, Pitoniak and Sievert framework for the small-x evolution \cite{Kovchegov:2015pbl,Kovchegov:2018znm,Cougoulic:2022gbk} a similar conclussion is reached.

The Electron-Ion Collider \cite{AbdulKhalek:2021gbh} will measure $g_1^p$ in the region of $10^{-5}<x<10^{-2}$ with unprecedented precision, exploring for the first time the behavior of $g_1^p$ and that of the gluon polarization in the small $x$ regime. In Figure 3 we show realistic pseudodata generated assuming a DSSV14 behavior but smeared according the expected experimental uncertainties for an accumulated integrated luminosity of 10 fb$^{-1}$ for center-of-mass system (c.m.s) energies of 45 and 140 GeV (open circles) \cite{Borsa:2020lsz}. On the other hand, we also show pseudodata produced from the Holographic-$A$ Pomeron prediction at the same energies (solid circles) assuming the experimental errors will be those computed in \cite{Borsa:2020lsz} for the corresponding kinematics. The pseudo data points are only those corresponding to a photon virtuality of $10$ GeV$^2$ for which the curves are computed. Even at plain sight it is clear that the EIC measurements will be able to discriminate between the two scenarios. For completeness, we have computed the impact of the future EIC measurements in both cases and we show it as new bands in darker blue and green for the DSSV and the Holographic-$A$ Pomeron scenarios, respectively. In the case of the Holographic-$A$ Pomeron it includes the original 56 experimental data plus 50 pseudo data points. The dark green error band is very narrow and cannot be discriminated from the corresponding central value since now the constant $C$ has the same central value as in equation (\ref{Cconstant}) but its error becomes
$7.64 \times 10^{-6}$, which means that the error is 200 times smaller than in (\ref{Cconstant}) where only the 56 experimental points were included. This behavior is due to the extremely high precision of the expected EIC measurements.

%---------------------------------------------------------------
%
\section{Conclusions}
%
%---------------------------------------------------------------

The string theory dual description of DIS and perturbative QCD offer complementary insights into phenomena that already are, or will be in the foreseeable future, probed by experiments with remarkable precision. In this paper we have confronted their respective predictions and the corresponding data to assess to which extent they overlap with good descriptions of the data, and where they complement each other. We have found an impressive agreement between the BPST Pomeron estimate for the unpolarized structure function $F_2^p$ and those coming from DGLAP based fits in a significant portion of the relevant kinematical range. This happens not only in the region covered by DIS data, where both approaches should agree by design, 
but also at low values of the parton momentum fraction $x$ and intermediate values of the photon virtuality $Q^2$, for which there is no data constraining the BPST Pomeron parameters. However, large discrepancies can be seen at higher values $Q^2$ where one expects the BPST Pomeron approach to be poorly constrained while DGLAP is in good standing. On the other hand, DGLAP estimates fail to agree between themselves and with the 
BPST Pomeron towards lower values of $Q^2$ where BPST Pomeron best reproduce the unpolarized DIS data. Of course at low $Q^2$
is where the convergence of the DGLAP perturbative series is weaker. This emphasizes the complementarity between both perspectives and gives a quantitative assessment of their respective limitations. 

In the case of the helicity-dependent structure function $g_1^p$ the available data is not as comprehensive as in the unpolarized case, but is enough to constrain the Holographic-$A$ Pomeron proposed in \cite{Kovensky:2018xxa}, and make a prediction for the forthcoming EIC experiment, that differs with the most standard DGLAP motivated predictions and suggest a significant role of the angular momentum in proton spin budget.

~

~

%%%%%%%%%%%%%%%%%%%%%%%%%%%%%%%%%%%%%%%%%%
%
\centerline{\large{\bf Acknowledgments}}
%
%%%%%%%%%%%%%%%%%%%%%%%%%%%%%%%%%%%%%%%%%%

~

The work of I.B., R.S. and M.S. has been supported in part by the Consejo Nacional de Investigaciones Cient\'{\i}fi\-cas y T\'ecnicas of Argentina (CONICET). The work of D.J. and M.S. has been supported in part by the Agencia Nacional para la Promoci\'on de la Ciencia y la Tecnolog\'{\i}a of Argentina (ANPCyT-FONCyT) Grant PICT-2017-1647, the UNLP Grant PID-X791, and the CONICET Grants PIP-UE B\'usqueda de nueva f\'{\i}sica and PICT-E 2018-0300 (BCIE).

\pagebreak

\newpage

\begin{appendices}
\section{A road map of soft, BFKL, BPST and Holographic-$A$ Pomerons for the uninitated.}

In the following, we disccuss with more detail what are the soft and hard Pomerons, and how 
these concepts developed in the context of the S-matrix and the gauge theory approaches, connect to the more recent gauge/string theory duality leading to the holographic dual 
description of the Pomeron physics at strong coupling.

%Before introducing the BPST and the Holographic-$A$ Pomerons, we will very briefly describe what are the soft and hard Pomerons. The idea is to make connexions between the previous $S$-matrix and gauge theory approaches and the more recent gauge/string theory duality perspective, which leads to the holographic dual description of the Pomeron physics at strong coupling. 

%---------------------------------------------------------------
%
\subsection{The soft Pomeron and the BFKL Pomeron}
%
%---------------------------------------------------------------

The dominant contribution to DIS at low $x$ comes from the gluon dynamics and the quark-antiquark sea. The standard DGLAP description at NLO in QCD should fail to describe the low-$x$ region since sub-leading terms in $\ln(Q^2/\mu^2)$ (where $\mu$ is an energy scale) involve powers of $\alpha_{\text {strong}} \ln(1/x)$, which become large (order 1) as $x \rightarrow 0$ \cite{Devenish:2004pb}. For example, let us suppose that the virtuality is $Q^2 \sim 10$ GeV$^2$ and $x \sim 10^{-2}$, thus $\alpha_{\text {strong}}$ is approximately 0.2, then $\alpha_{\text {strong}} \ln(1/x) \approx 0.4$.

During the sixties, strong interactions were investigated within the $S$-matrix formalism, leading to the Regge theory, used to calculate hadron-hadron and photon-hadron cross-sections at high energy \cite{Devenish:2004pb}.
The $S$-matrix elements between two asymptotic states, one in the remote past and another in the remote future, is given by $S_{i \, f} = \langle f|\hat{S}|i \rangle= \langle f_{out}| i_{in} \rangle$. There are three very important postulates about the $S$ matrix, namely: Lorentz invariance, unitarity and analyticity \cite{Forshaw:1997dc,Donnachie:2002en}.

Let us consider a two-to-two particles scattering, with incoming particles $i_1$ and $i_2$ and the outgoing ones $f_3$ and $f_4$. The incoming four-momenta are $p_1^\mu$ and $p_2^\mu$ and the outgoing four-momenta are $p_3^\mu$ and $p_4^\mu$, while their masses are $m_j$ ($j=1, \cdots, 4 $), respectively. This process can be described in terms of the Mandelstam variables defined in equations (\ref{stu}). Recall that $t$ is the square of the four-momentum exchanged between particles $i_1$ and $f_3$, and there is also the kinematic relation among them, which obviously implies that the transition amplitude for the process $i_1+i_2 \rightarrow f_3 + f_4$ is a function of only two independent Mandelstam variables, ${\cal {A}}(s, t)$.

Unitarity of the $S$ matrix, $\hat{S} \hat{S}^\dagger = \hat{S}^\dagger \hat{S} =\mathbb{I}$, implies that the probability for the transition between incoming and outgoing states when all possible final states are added is one. Thus, 
\begin{equation}
S_{i \, f} = \delta_{i \, f} + i \, (2 \pi)^4 \, \delta^{(4)}\left(\sum_{i} p_i - \sum_{f} p_f\right) \, {\cal {A}}_{i \, f} = \delta_{i \, f} + i \, T_{i \, f} \, . \label{Sif}
\end{equation}
For a two-to-two particle scattering, being $|a\rangle$ a two-particle state, using equation (\ref{Sif}) and the unitarity condition, it leads to the optical theorem,
\begin{equation}
2 \, \text{Im} {\cal {A}}_{a \, a} = (2 \pi)^4 \, \sum_X \, \delta^{(4)}\left(\sum_{a} p_a - \sum_{f} p_f\right) \,  {\cal |{A}}_{a \rightarrow X}|^2 \propto \sigma_{\text{Total}}\, ,
\end{equation}
where $X$ represents intermediate states. $\sigma_{\text{Total}}$ is the total cross-section for the scattering. When the center-of-mass energy is much larger that the masses of the incoming particles it leads to $\text{Im} {\cal {A}}_{a \, a}  \sim  s \, \sigma_{\text{Total}}$.

In addition, analyticity implies that the $S$ matrix is an analytic function of the Lorentz invariants, and it only has the singularities allowed by unitarity. Also, analyticity implies crossing symmetry: ${\cal {A}}(s, t) = {\cal {A}}(t, s)$. From analyticity and unitarity one can extract the $s$-plane singularity structure. In particular, considering the $t$-channel and the high energy limit $|t| \ll s$, the amplitude can be expanded in terms of Legendre polynomials $P_l(\cos \theta)$, where $\theta$ is the scattering angle in the center-of-mass frame, which can be written as $\cos \theta = 1+2t/s$. It leads to the partial wave expansion, which after using the crossing symmetry ($s \leftrightarrow t$), becomes
\begin{equation}
{\cal {A}}(s, t) = \sum_{l=0}^\infty (2 l+1)\ a_l(t) \ P_l(1+2 s/t) \,  , \label{AmplitudeLegendrePols}
\end{equation}
with the partial wave amplitudes  $a_l(t)$ .

At this point it is instructive to recall what happens if a single resonance with mass $M_J$ and spin $J$ gives the leading contribution to the $t$-channel process. In this case the high energy behavior of the corresponding amplitude is
\begin{equation}
{\cal {A}}(s, t) \sim  \frac{C}{t-M_J^2} \ \left(\frac{2 s}{t}\right)^J \,  , \label{AmplitudeJ}
\end{equation}
which obviously becomes very large in the high energy limit $s \gg |t|$, in fact unbound, which indicates that a single resonance exchange in the $t$-channel cannot be the leading contribution. On the other hand, the amplitude (\ref{AmplitudeLegendrePols}) can be rewritten in terms of a contour integral in the complex plane of the angular momentum $l$ \cite{Forshaw:1997dc}
\begin{equation}
{\cal {A}}(s, t) = \frac{1}{2 i}  \oint_C dl \ (2 l+1) \ \frac{a(l, t)}{\sin(\pi l)} \ P(l, 1+2s/t) \,  ,
\end{equation}
where the contour $C$ surrounds the positive real axis.
Notice that $a(l, t)$ is an analytic continuation of $a_l(t)$ in (\ref{AmplitudeLegendrePols}). In order to be more precise, the analytic structure of the function $a(l, t)$ requires two analytic continuations corresponding to the even and odd partial wave amplitudes $a^{(\eta_{\text{sign}})}(l, t)$, with $\eta_{\text{sign}} = \pm 1$.
Next, one has to deform the contour $C$ to another contour parallel to the imaginary axis located at $\text{Re} \ l = -1/2$, and encircling any poles or cuts that the functions $a^{(\eta_{\text{sign}})}(l, t)$ may have at $l=\alpha_{\eta_{\text{sign}}}(t)$, which are the Regge poles. Then, in the $|t| \ll s$ limit the scattering amplitude becomes
\begin{equation}
{\cal {A}}(s, t) \rightarrow \left( \frac{\eta_{\text{sign}}+\exp[-i \pi \alpha(t)]}{2}\right) \ \beta(t) \ s^{\alpha(t)} \,  , \label{AmplitudeRegge}
\end{equation}
where $\alpha(t)$ represents the leading Regge trajectory, while the function $\beta(t)$ contains the residues of the poles of the complex angular momentum integral multiplied by other factors.

The comparison of amplitudes (\ref{AmplitudeRegge}) and (\ref{AmplitudeJ}) suggests that the former can be understood as  the one generated by the exchange of an object carrying angular momentum $\alpha(t)$, called Reggeon, which is not a single particle. Therefore, the amplitude (\ref{AmplitudeRegge}) can be interpreted as the superposition of amplitudes corresponding to the exchanges of all possible particles in the $t$-channel, which leads to a Regge trajectory. For $t>0$ it is expected to have the poles corresponding to the exchange of particles of spin $J=\alpha(M_J^2)$ and mass $M_J$. Chew-Fraustchi plot suggests that there is a linear Regge trajectory
\begin{equation}
\alpha(t) = \alpha_1 \ t + \alpha_0 \, ,
\end{equation}
where $\alpha_1$ is the Regge slope and $\alpha(0)=\alpha_0$ is the intercept. From the high-energy limit of the total cross-section
\begin{equation}
\sigma_{\text{Total}} \rightarrow s^{\alpha(0)-1} \, , \label{alpha-Pomeron}
\end{equation}
it can be extracted the leading Regge trajectory. Experimental data show that $\sigma_{\text{Total}}$ increases slowly with $s$. Assuming that this increase is induced by the exchange of a Reggeon, its intercept must be larger than one. Moreover, the amplitude must be dominated by the exchange of a Reggeon with zero isospin and it has to be even under charge conjugation. This particular Reggeon is called the soft Pomeron (recall that this is for positive values of $t$), and it is postulated to be a bound state of gluons referred as glueball. The Pomeron intercept has been obtained from the fit of equation (\ref{alpha-Pomeron}) to the proton-proton cross-section experimental data, obtaining to $\alpha_P(0)=1.0808$ \cite{Donnachie:1992ny}.

The relation to the symmetric structure functions $F_1$ and $F_2$ comes from the fact that DIS can be written in terms of the $\gamma^* p$ scattering process, where $\gamma^*$ represents a virtual photon, with squared four-momentum $q^2=-Q^2$, 
\begin{equation}
\sigma_{\text {Total}}^{\gamma^* p}(W^2, Q^2) = \sigma_T + \sigma_L \approx \frac{4 \pi^2 \alpha_{em}}{Q^2} \ F_2(x, Q^2) \, , \label{sigmaTotal}
\end{equation}
where $\alpha_{em}$ is the fine structure constant and (for $W^2$ much larger than the square of the proton mass) it leads to $W^2=Q^2 (1/x-1)$, which at low $x$ becomes $W ^2 \approx Q^2/x$, where $W$ is the center-of-mass energy of the $\gamma^*+p$ system.
The low-$x$ the behavior of $\sigma_{\text {Total}}^{\gamma^* p}(W^2, Q^2)$ in equation (\ref{sigmaTotal}) is dominated by the exchange of a single Pomeron, leading to $F_2(x, Q^2) \propto x^{-0.08}$ as $x$ becomes very small.

~

There is another Pomeron, called hard or BFKL Pomeron, and in the rest of this subsection we briefly describe it. In the context of Regge theory a given particle of mass $M$ and spin $J$ is said to Reggeize if the scattering amplitude corresponding to a process, that in its $t$-channel exchanges the quantum numbers of that particle, goes like ${\cal {A}}(s, t) \propto s^{\alpha(t)}$. We identify the Regge trajectory $\alpha(t)$ while the spin and mass follow the relation $\alpha(M_J^2)=J$, being the particle on Regge trajectory. The lowest order Feynman diagram from QCD which perturbatively can simulate a Pomeron exchange like this is given by a two-gluon exchange. In fact the Pomeron in QCD is constructed from ladder diagrams whose vertical lines are Reggeized gluons. The ladders are completed with rungs connected to the vertical Reggeized gluons through effective vertices. In this context the behavior of this Pomeron is derived from the BFKL equation \cite{Fadin:1975cb,Kuraev:1976ge,Kuraev:1977fs,Balitsky:1978ic}. Notice that there is an infinite sum of the described ladder Feynman diagrams with different number of rungs, and there is no color exchange through the vertical lines. A diagram with $n$ rungs contributes with a factor $(\alpha_s \log s)^n$. In this appendix we use the traditional notation for the QCD coupling $\alpha_s$ instead of $\alpha_{\text{strong}}$ that we use in the main text.

Let us consider the most general Feyman diagram consisting in the exchange of two Reggeized gluons (vertical ladders) between two quarks (horizontal lines at the top and the bottom of the diagram). This type of diagram corresponds to a quark-quark scattering, and it would be related to the DIS diagram with the exchange of a BFKL Pomeron. The BFKL amplitude $\hat f(\omega, k_1, k_2, q)$, where $k_1$ and $k_2$ are the transverse momenta with which the quark in the top line and the quark in the bottom horizontal line are probed by the BFKL Pomeron, respectively, while $q$ is the momentum transfer. The color singlet ($S$) quark-quark (no color exchange in the $t$ channel) scattering amplitude is given by
\begin{equation}
\tilde{\cal {A}}_{qq}^S(\omega, t) = \int_1^\infty dz \ z^{-\omega-1} \ \frac{{\cal {A}}_{qq}^S(s, t)}{s} =
4 i \alpha_s^2 G^S \int \frac{d^2k_1 d^2k_2}{k_2^2 (k_1-q)^2}  \hat f(\omega, k_1, k_2, q) \, ,
\end{equation}
where $\tilde{\cal {A}}_{qq}^S(\omega, t)$ is the Mellin transform, while $z=s/k^2$ with $k^2$ being a scale factor related to the external transverse momenta. $G^S$ projects out the color singlet term. The total cross-section needs only the forward amplitude, namely ${\cal {A}}_{qq}^S(s, t=0)$, therefore one needs only $\hat f(\omega, k_1, k_2, q=0)$. It is convenient to consider the inverse Mellin transform of $\hat f(\omega, k_1, k_2, q=0)$ that we call $F(s, k_1, k_2)$. By considering only the leading ($n=0$) term, the function $F(s, k_1, k_2)$ is given by the following integral in the complex plane $\gamma$
\begin{equation}
F(s, k_1, k_2) = \int \frac{d\gamma}{2 \pi i} \ \left(\frac{s}{k^2}\right)^{\bar{\alpha}_s \chi}  \frac{1}{\pi k_1^2} \left(\frac{k_1^2}{k_2^2}\right)^\gamma \, ,
\end{equation}
where $\bar{\alpha}_s=3 \alpha_s/\pi$ and the contour runs parallel to the imaginary $\gamma$ axis. The function $\chi(\gamma)$ is given by $\chi(\gamma)= 2 \psi(1)-\psi(\gamma)-\psi(1-\gamma)$, with $\psi(x)=d\log(\Gamma(x))/dx$, with the usual Gamma function. 
The leading $s$ behavior leads to
\begin{equation}
F(s, k_1, k_2) = \frac{1}{\pi k_1 k_2} \left(2 \pi \bar{\alpha}_s |\chi''(1/2)|\log(s/k^2)\right)^{-1/2} \left(\frac{s}{k^2}\right)^{\bar{\alpha}_s \chi(1/2)}   \, ,
\end{equation}
which leads to the high-energy behavior of the quark-quark scattering amplitude
\begin{equation}
\tilde{\cal {A}}_{qq}^S(s, t=0) \sim \frac{(s/k^2)^{1+\omega_0}}{\sqrt{\log{(s/k^2)}}}  \, ,
\end{equation}
with $\omega_0=\bar{\alpha}_s \chi(1/2)=4 \bar{\alpha}_s \log 2$. Typically it leads to a very strong rise of the quark-quark total cross-section in the high-energy limit.

Now, we focus on the application of the BFKL formalism to DIS.  Using the optical theorem we take the imaginary part of the elastic $\gamma^* + \text{proton} \rightarrow \gamma^* + \text{proton}$ cross-section at $t=0$ we can use the BFKL function $F(W^2, k_1, k_2)$ convoluted with the proton and the photon impact factors $\Phi_p(k_2)$ and $\Phi_{\gamma,  \epsilon}(k_1)$, respectively, 
\begin{equation}
\sigma_\epsilon^{\gamma^* p}(W^2, Q^2)= \frac{1}{(2 \pi)^4} \int \frac{d^2 k_1}{k_1^2} \int \frac{d^2 k_2}{k_2^2} \ \Phi_{\gamma,  \epsilon}(k_1) \ \Phi_p(k_2) \ F(W^2, k_1, k_2) \, , \label{sigmas}
\end{equation}
where 
\begin{equation}
W^2=(p+q)^2  \, ,
\end{equation}
is the the square of the center-of-mass energy of the virtual photon-proton system. $p_\mu$ and $q_\mu$ are the four-momenta of the proton and the virtual photon, respectively. Recall that $Q^2=-q^2>0$. The polarization $\epsilon$ can be transverse (T) or longitudinal (L). For inclusive DIS the proton impact factor cannot be calculated in perturbation theory. For the DIS of an electron with four-momentum $k_\mu$ off a proton of four-momentum $p_\mu$ we can define the following kinematic variables
\begin{eqnarray}
s&=&(p+k)^2 \, \\
x&=&\frac{Q^2}{2 p \cdot q} \approx \frac{Q^2}{Q^2+W^2} \, , \\
y&=& \frac{p \cdot q}{p \cdot k} \approx \frac{Q^2}{x s} \, ,
\end{eqnarray}
where $x$ is the Bjorken variable. Assuming the limit where the electron and the proton masses are negligible compared with the energy scale of DIS the approximate equalities become exact. We also assume that $W^2 \gg Q^2 \gg M_p^2$, from which it follows that $0<x\ll 1$. The proton structure function $F_2^p$ and the longitudinal one $F_L^p$ are related to the photon-proton cross-sections (\ref{sigmas}) by
\begin{eqnarray}
F_2(x, Q^2) &=& \frac{Q^2}{4 \pi^2 \alpha_{em}} \left(\sigma^{\gamma^* p}_T(x, Q^2) + \sigma^{\gamma^* p}_L(x, Q^2) \right)\, , \label{F2sigmaT} \\
F_1(x, Q^2) &=& \frac{Q^2}{4 \pi^2 \alpha_{em}} \sigma^{\gamma^* p}_L(x, Q^2)\, ,
\end{eqnarray}
The first of these equations is similar to equation (\ref{sigmaTotal}). The proton impact factor cannot be obtained from perturbation theory in QCD for obvious reasons, therefore it must be modelled. There is another important issue due to the fact that in QCD the next-order correction to the BFKL Pomeron is large, and it comes with opposite sign with respect to the single BFKL Pomeron itself \cite{Fadin:1998py,Camici:1997ij}. We should emphasize that the BFKL Pomeron is derived from a perturbative calculation in QCD.

%---------------------------------------------------------------
%
\subsection{The BPST Pomeron and the unpolarized function $F_2^p$}
%
%---------------------------------------------------------------

In certain hadron scattering processes at high energy ($s \gg |t| \gg \Lambda^2_{QCD}$, where $\Lambda_{QCD}$ is the IR scale of QCD) and small scattering angle the Regge theory suggests the exchange of a soft Pomeron (Reggeon) for positive $t$ values, and a single BFKL-Pomeron exchange at leading order in $\alpha_s \log s$ at weakly coupled QCD for $t \leq 0$. The soft Pomeron is understood as an exchange of a single glueball, which in the string theory dual language it corresponds to a closed string. Besides, the BFKL Pomeron represents the exchange of a color-singlet object composed by Reggeized gluons. It is worth to consider the large $N_c$ limit a gauge theory\footnote{$N_c$ is the rank of the gauge group $SU(N_c)$.} since in that case many aspects of the gauge theory becomes simpler, both to calculate and interpret. This limit is not real ($N_c=3$) QCD but a related gauge theory. 
This limit leads to that a single Pomeron exchange dominates the scattering amplitude. Otherwise, for finite $N_c$ one may expect multiple-Pomeron exchanges become important and eventually may dominate the high-energy behavior of scattering amplitudes. From the type IIB superstring theory perspective the dual exchanged object is a Reggeized graviton, leading to the BPST Pomeron \cite{Brower:2006ea}. Thus, for strongly coupled gauge theory the BPST Pomeron unifies the soft and hard Pomerons, something which technically is not possible in QCD. This property of the BPST Pomeron is very important.

Let us describe very briefly the ideas behind the derivation of the BPST Pomeron from type IIB superstring theory \cite{Brower:2006ea}. A dual representation of a hard scattering process of two hadrons to two hadrons at high energy may be described in terms of a four-point superstring theory scattering amplitude \cite{Polchinski:2001tt}. In particular, in the Regge limit and at strong coupling of the gauge theory, the dual description leads to a BPST Pomeron exchange. Strictly speaking, the holographic dual calculation is valid for $N_c \gg \lambda_{\text{'t Hooft}} \gg 1$. Therefore, within the framework of perturbation theory of superstring theory, which means that the string theory coupling $0<g_{\text{string}} \ll 1$, one only needs to consider a world-sheet given by a two-dimensional sphere represented by coordinates $(\sigma_1, \sigma_2)$. The closed string proper time is $\sigma_1$ and its proper length is $0 \leq \sigma_2 \leq 2 \pi$. The ten-dimensional ambient space where the closed string propagates is described by fields which take values on the string world-sheet 
\begin{equation}
X^M(\sigma_1, \sigma_2) = x^M + X'^M(\sigma_1, \sigma_2) \, , \label{coordinates}
\end{equation}
where $x^M$ with $M=0, \dots 9$ labels the closed string center-of-mass position, and $X'^M(\sigma_1, \sigma_2)$ characterize the string vibrations. By taking $x^M$ constant, the Gaussian integral on $X'^M$ (which is needed for the quantization of the string theory) leads to exactly the same result as it would do in ten-dimensional Minkowski space-time. This gives the ten-dimensional flat-space $S$ matrix that would be seen by a local observer,
\begin{equation}
S = i \int d^4x \ \int d^6y \ \sqrt{-G} \ A_{\text{local}}(x, y) \, .
\end{equation}
$x_\mu$ denotes coordinates in the four-dimensional Minkowski space-time and $y_\alpha$ is used for the radial and the angular coordinates on the five-dimensional sphere $S^5$. This is a local approximation which allows to carry out the calculations by replacing the (unknown) superstring theory scattering amplitude of the curved AdS$_5 \times S^5$ space-time by the (known) superstring theory scattering amplitude in ten-dimensional Minkowski space-time, $A_{\text{local}}(x, y)$. For instance, if one considers the Regge limit of the proton+proton to proton+proton scattering amplitude in the gauge theory at strong coupling, the corresponding superstring theory scattering amplitude $A_{\text{local}}(x, y)$ of the dual type IIB superstring theory description is the flat ten-dimensional scattering amplitude of four-closed strings. Each of these closed strings can be viewed as the insertion of a dilatino vertex operator on the two-dimensional spherical world-sheet. This is the starting point of the calculation. Then, the amplitude can be expressed as follows
\begin{equation}
 A_{\text{local}}(x, y) \rightarrow \tau_{10}(\tilde{P}) \ \prod_{i=1}^4 \ e^{i p_i \cdot x_i} \ \Psi(y_i) \, ,
\end{equation}
where $\tau_{10}(\tilde{P})$ is the ten-dimensional flat space-time string theory scattering amplitude, which only depends on the momenta $\tilde{P}$ seen by a local inertial observer in the AdS$_5 \times S^5$ bulk. In addition, the four external states are represented by four free wave-functions in the flat four-dimensional Mikowski space-time times the corresponding wave-functions depending of the internal coordinates $\Psi(y_i)$. 

Then, one can obtain the $S$-matrix given by
\begin{equation}
S = i \ (2\pi)^4 \ \delta^{(4)}\left(\sum_{i=1}^4 p_i \right) \ \int d^6y \ \sqrt{-G_{6d}} \ \tau_{10}(\tilde{P}) \ \prod_{i=1}^4 \Psi(y_i) \, ,
\end{equation}
where $G_{6d}$ is the determinant of the part of the metric which contains the radial coordinate $r$ and the five sphere. $\delta^{(4)}\left(\sum_{i=1}^4 p_i \right)$ comes from the four-dimensional integral and ensures the conservation of the four-momentum.

We should emphasize that due to the metric warp factor there is a red-shift as mentioned before
\begin{equation}
\tilde{P}^\mu_{\text{10d}} = \frac{R}{r} \ p^\mu_{\text{4d}}  \, \nonumber
\end{equation}
being $\tilde{P}^\mu_{\text{10d}}$ the inertial four-momentum measured by a local observer in the bulk, while $p^\mu_{\text{4d}}$ is the same component of the four-momentum corresponding to the gauge theory at the boundary of the AdS space. The metric warp factor also induces the red-shift of the Mandelstam variables 
\begin{equation}
\tilde{s}_{\text {10d}} = \frac{R^2}{r^2} \ s \,\,\,\,\,\,\,\,\,\, \text{and} \ \ \,\,\,\,\,\,\,\,
\tilde{t}_{\text{10d}} = \frac{R^2}{r^2} \ t  \, .
\end{equation}
Notice that we have dropped the 4d sub-indices of the four-dimensional Mandelstam variables. From superstring theory we have
\begin{equation}
\tau_{10}(\tilde{P}) = g_{\text{string}}^2 \ \alpha'^3 \ F_s(\tilde{P} \sqrt{\alpha'}) \, ,
\end{equation}
where the function $F_s$ is given by
\begin{equation}
F_s(\tilde{P} \sqrt{\alpha'}) = K(\tilde{P} \sqrt{\alpha'}) \ \left[
\prod_{\tilde{x}=\tilde{s},\tilde{t},\tilde{u}} \
\frac{\Gamma(-\alpha' \tilde{x}/4)}{\Gamma(1+\alpha' \tilde{x}/4))} \right] 
\, ,
\end{equation}
which, for $|\tilde{t}| \ll \tilde{s}$, with $\tilde s+\tilde t+\tilde u=0$, can be approximated by
\begin{equation}
F_s(\tilde{P} \sqrt{\alpha'})   \approx K(\tilde{P} \sqrt{\alpha'}) \  \frac{\Gamma(-\alpha' \tilde{t}/4)}{\Gamma(1+\alpha' \tilde{t}))} \ (\alpha'  \tilde{s})^{2+\alpha' \tilde{t}/2} = f(\alpha' \tilde{t}) \ (\alpha'  \tilde{s})^{2+\alpha' \tilde{t}/2} \, ,
\end{equation}
where in order to abbreviate the notation we have dropped the sub-index 10d in the ten-dimensional Mandelstam variables. $K(\tilde{P} \sqrt{\alpha'})$ is a kinematic factor.
Using these expressions in $\tau_{10}(\tilde{P})$ we obtain the four-dimensional scattering amplitude (depending on the four-dimensional Mandelstam variables $s$ and $t$)
\begin{equation}
\tau_{4}(s, t)
 = \int d^6y \ \sqrt{-G} \ \Psi_3(y) \ \Psi_4(y) \ f(\alpha' \tilde{t}) \ (\alpha'  \tilde{s})^{2+\alpha' \tilde{t}/2} \  \Psi_1(y) \ \Psi_2(y)\, . \label{tau1}
\end{equation}
The relevant exponent in the Regge limit is $j=2+\alpha' \ \tilde{t}/2=2+\alpha' \ t \ R^2/(2 r^2)$, which is a very important result as we show in what follows. 

Let us show that this expression leads to two very different physical situations. Firstly, let us consider the case of positive $t$ and the Regge limit $0<t \ll s$, for which the maximum value of the exponent is reached when the radial coordinate $r$ has its minimum $r_0$, and since $r_0 \leq r $, it corresponds to the IR of the gauge theory. Therefore, it is  related to the soft Pomeron at strong coupling. Thus, we obtain
\begin{equation}
j_{\text{Max}}=2+\alpha' \tilde{t}/2=2+\alpha' t R^2/(2 r_0^2) \, , \label{intercept2}
\end{equation}
which shows a linear Regge trajectory $j_{\text{Max}}(t)$, with intercept $\alpha_0=2$ and slope $\alpha_1=\alpha' R^2/(2 r_0^2)$. 

The second possibility in the study of the exponent corresponds to $t<0$ and $0<|t| \ll s$ where the maximum value of the exponent is 
\begin{equation}
j_{\text{Max}}=2 \, ,
\end{equation}
which corresponds to $r \rightarrow \infty$, namely the UV region of the gauge theory, related to the BFKL Pomeron. This is the effect of unification (or interpolation) of the soft and the BFKL Pomerons that we mentioned before \cite{Brower:2006ea}.

The derivation presented so far deals with a local approximation, which implies to consider the large $\lambda_{\text{'t Hooft}}$ limit, leading to the Gaussian approximation (see discussion below equation (\ref{coordinates})), and then the high energy limit. However, in order to reach a more realistic parametric domain of QCD, it is crucial to investigate the physics for values of $s$ growing as $\exp(\lambda_{\text{'t Hooft}}^{1/2})$. From the gauge/string theory duality this is an extremely large energy scale, however, in order to consider QCD one has to explore what happens towards smaller values of the 't Hooft coupling, which means realistic values of the QCD coupling. As we have seen in this work, this leads to a very precise description of the proton structure functions when this formalism is applied to DIS. Thus, we must retain terms of order $\lambda_{\lambda_\text{'t Hooft}}^{-1/2}$ in the exponent $j=2+\alpha´\ \tilde{t}/2$ in the scattering amplitude (\ref{tau1}), which implies to consider the ten-dimensional momentum operator ($i \partial^\mu_{10}$) in the definition of $\tilde{t}$,
\begin{equation}
\alpha' \tilde{t} \rightarrow \alpha' \nabla_P^2 \equiv \alpha' \frac{R^2}{r^2} \ t + \alpha' \nabla_\perp^2 \, , \label{nablaP}
\end{equation}
where $\nabla_\perp^2$ is the Laplacian operator in the radial and five-dimensional angular directions, which is proportional to $\alpha'/R^2 = \lambda_{\text{'t Hooft}}^{-1/2}$, and acts on the wave-functions of the incoming and outgoing states. The transverse momentum transfer leads to the ${\cal {O}}(\lambda_{\text{'t Hooft}}^{-1/2})$ correction to the intercept as shown in equation (\ref{intercept2}), and also it makes $s^{\alpha' \tilde{t}/2}$ a diffusion operator in the eight transverse directions\footnote{From the ten dimensions of type IIB superstring theory there are two directions defining the so-called light-cone coordinates, the time and the direction of motion of the two head-on colliding closed strings, the eight remaining ones are the transverse coordinates.}, which induces a diffusion operator similar to the one corresponding to the BFKL Pomeron. In addition, it will show important changes in comparison with the local approximation where the second term in (\ref{nablaP}) was ignored.

Now, let us write the Laplacian $\nabla^2_\perp$, considering the metric (\ref{metric1}). The idea is to include the contribution of the $t$-channel exchange of a generic transverse traceless tensor field of spin $j$, $\Phi_{+j} \equiv \Phi_{++ \dots +}$, with $j$ light-cone indices $+$, being $x^\pm = (x^0 \pm x^1)/\sqrt{2}$ the light-cone coordinates. This represents a fluctuation of a generic field propagating in the AdS$_5 \times S^5$ bulk. In particular, in the case of the BPST Pomeron it corresponds to $j=2$ and it is given by transverse traceless fluctuations of the metric. On the other hand, in the case of the Holographic-$A$ Pomeron it corresponds to $j=1$ and the fluctuation is given by a linear combination of the gravi-photon and the Ramond-Ramond four-form field $A_4$ in type IIB superstring theory. Thus, the covariant Laplacian acting on a $\Phi_{++}$ is given by 
\begin{equation}
\nabla^2_2 \Phi_{++} = \frac{r^2}{R^2} \nabla_0^2 \left(\left(\frac{R^2}{r^2}\right) \Phi_{++} \right) + \frac{1}{2} {\cal{R}}^{\ +}_+ \, ,
\end{equation}
where ${\cal{R}}^{\ +}_+$ is the Ricci tensor $++$ components, and $\nabla_0^2$ is the scalar Laplacian ($j=0$). From the equations of motion of type IIB supergravity one obtains
\begin{equation}
\Delta_2 \Phi_{++} \equiv \frac{r^2}{R^2} \nabla_0^2 \left(\left(\frac{R^2}{r^2}\right) \Phi_{++} \right) = 0 \, ,
\end{equation}
when $\Phi_{++}$ is a transverse traceless fluctuation of the metric. Then, the $\lambda_{\text{'t Hooft}}^{-1/2}$ correction to the exponent in the amplitude (\ref{tau1}) leads to
\begin{equation}
\tau_{4}(s, t)
 = \int d^6y \ \sqrt{-G} \ \Psi_3(y) \ \Psi_4(y) \ f(\alpha' t R^2/r^2) \ (\alpha'  s R^2/r^2)^{2+\alpha' \Delta_2/2} \  \Psi_1(y) \ \Psi_2(y)\, . \label{tauNabla}
\end{equation}
In order to calculate this amplitude at high energy it is convenient to make a change of coordinates in the metric (\ref{metric1}) given by $u = \log(r/r_0)$, which at large $r$ reads
\begin{equation}
ds^2= \frac{r_0^2}{R^2} \, e^{2 u} \, \eta_{\mu\nu} dx^\mu dx^\nu + \frac{R^2}{r_0^2} dr^2 + R^2 d\Omega_5^2 \, . \label{metricU}
\end{equation}
Notice the presence of the additional warp factor $e^{2 u}$ in front of the first piece of this metric.
Then, one may calculate the imaginary part of the scattering amplitude (recall that this is related to the cross-section of the process), leading to 
\begin{equation}
{\text{Im}} {\cal {A}}(s, t=0)
 \propto \int du \int du' \ \Psi_3(u) \ \Psi_4(u) \ {\cal {K}}(u, u', \tau_b, t=0) \  \Psi_1(u') \ \Psi_2(u')\, , \label{tauNablaU}
\end{equation}
where the BPST kernel is given by ${\cal {K}}(u, u', \tau_b, t=0) = s^{j_0} \ {\cal {K}}_0(u, u', \tau_b, t=0)$, being
\begin{equation}
j_0 =2 - \frac{2}{\lambda_{\text{t' Hooft}}^{1/2}} \, ,
\end{equation}
which can be identified with the strong coupling limit of the BFKL Pomeron exponent \cite{Brower:2006ea}. In addition we have
\begin{equation}
{\cal {K}}_0(u, u', \tau_b, t=0)= \frac{e^{-(u-u')^2/4 \tau_b}}{2 \sqrt{\pi \tau_b}} + {\cal {F}}(u, u', \tau_b)  \frac{e^{-(u+u')^2/4 \tau_b}}{2 \sqrt{\pi \tau_b}} \, ,
\end{equation}
where 
\begin{equation}
{\cal {F}}(u, u', \tau_b) = 1- 4 \sqrt{\pi \tau_b} e^{\eta^2} {\text{erfc}}(\eta) \, ,
\end{equation}
while 
\begin{equation}
\eta = \frac{u+u'+4 \tau_b}{\sqrt{4 \tau_b}} \, ,
\end{equation}
and
\begin{equation}
{\text{erfc}}(\eta) = \frac{2}{\sqrt{\pi}} \int_\eta^\infty dk \ e^{-k^2} \, .
\end{equation}
and $\tau_b$ is given by
\begin{equation}
\tau_b = \frac{1}{2 \lambda_{\text{' Hooft}}^{1/2}} \log\left(  \frac{R^2}{r^2} \alpha' s \right) \, .
\end{equation}
By increasing the center-of-mass energy $\sqrt{s}$, the exchange of multiple Pomerons is not suppressed and one must include them. There is a way to resume multiple Pomeron exchange known as the eikonal method \cite{Brower:2007qh,Brower:2007xg}. It implies to write the scattering amplitude in terms of the impact parameter $\vec b$. Thus, for a two-to-two on-shell hadrons scattering the amplitude can be written in an eikonal sum leading to
\begin{equation}
{\cal {A}}(s, t) = 2 i s \int d^2b \ e^{i {\vec q} \cdot {\vec b}} \ \int dr \int dr' \ P_{13}(r) \left(1-e^{i \chi_{\text{eikonal}}(s, b, r, r')} \right) P_{24}(r') \, ,
\end{equation}
where the eikonal is related to the BPST Pomeron kernel by
\begin{equation}
 \chi_{\text{eikonal}}(s, b, r, r') = \frac{g_0^2}{2 s} \left(\frac{r r'}{R^2}\right)^2 \ {\cal {K}}(s, b, r, r') \, ,
\end{equation}
$g_0^2$ is a parameter to be determined by fitting to experimental data, while we have expressed the BPST Pomeron kernel in terms of the variables $s, b, r$ and $r'$. 
$P_{13}(r)$ and $P_{24}$ label the impact factors associated to the scattered hadrons.

Now, let us focus on the DIS of an electron from a proton. The structure function $F_2$ can be calculated from the total cross-section corresponding to the off-shell photon-proton scattering, which by using the optical theorem, is proportional to the imaginary part of the forward off-shell amplitudes of $\gamma^*$+proton amplitude, $\sigma^{\gamma^* p}_{\text {Total}}={\text {Im}}{\cal {A}}(s, t=0)/s$ (see equation (\ref{F2sigmaT})). 
$F_2$ was derived from the BPST Pomeron in \cite{Brower:2010wf}. It has four free parameters: $g_0^{2}$, $\rho$, $z_0$ and $Q'$, obtained by fitting it to experimental data. Then
\begin{equation}
F^{\text{BPST}_{\text{HW}}}_2(x, Q^2) = \frac{g_0^2 \ \rho^{3/2} \ Q}{32 \ \pi^{5/2} \ \tau_b^{1/2} \ Q'} \ e^{(1-\rho) \tau_b} \left(e^{-\frac{\log^2{(Q/Q')}}{\rho \tau_b}}+ {\cal{F}}(x, Q, Q') \ e^{-\frac{\log^2{(Q Q' z^2_0)}}{\rho \tau_b}}\right) \, ,
\end{equation}
the supra-index HW indicates that this expression has been derived considering the IR hard-wall cut-off in the metric $r_0=R^2/z_0$. Also, we have 
\begin{eqnarray}
{\cal{F}}(x, Q, Q')= 1-2 \ (\pi \ \rho \ \tau_b)^{1/2} \ e^{\eta^2(x, Q, Q')} \ {\text{erfc}}\left(\eta(x, Q, Q')\right) \, ,
\end{eqnarray}
and
\begin{eqnarray}
\eta(x, Q, Q') &=& \frac{\log{\left(Q'\ Q \ z_0^2 \right)}+ \rho \ \tau_b}{\sqrt{ \rho \ \tau_b}} \ , 
\end{eqnarray}
where
\begin{eqnarray}
\tau_b(x, Q, Q') &=& \log{ \left( \frac{\rho \ Q }{2 Q' x} \right)} \, , 
\end{eqnarray}
is a longitudinal boost. 

The parameter $Q'$ is approximately $r'/R^2$, being $r'/R^2$ the support of the Dirac's delta function used to approximate the hadron impact factor \cite{Brower:2010wf}. Therefore, $r'$ should be of the order of the hadron size and $Q'$ must be of the order of the proton mass. In addition, the virtual-photon impact factor is also approximated by a Dirac's delta function peaked at $Q \approx r/R^2$. The parameter $\rho$ is related to the 't Hooft coupling $\rho=2/\lambda_{\text{t' Hooft}}^{1/2}$, and $z_0 \equiv R^2/r_0$ is the IR cut-off of the gauge theory ($\Lambda \equiv r_0/R^2$). Thus, there is a clear physical interpretation of these parameters.

%---------------------------------------------------------------
%
\subsection{The Holographic-$A$ Pomeron and the polarized function $g_1^p$}
%
%---------------------------------------------------------------

In order to study the $g_1$ helicity function let us firstly very briefly discuss where it comes from, by considering the DIS differential cross-section corresponding to polarized charged leptons scattered off polarized hadrons. We consider a final polarized lepton in the solid angle $d \Omega$ and in the final energy range $(E', E'+dE')$ 
\begin{equation}
\frac{d^2 \sigma}{d \Omega \ dE'} =  \frac{\alpha_{em}^2}{2 M q^4} \ \frac{E'}{E} \ l_{\mu\nu} \ W^{\mu\nu} \, ,
\end{equation}
in the laboratory frame \cite{Anselmino:1994gn}. Thus, the hadron four-momentum is $P_\mu = (M, 0)$ of mass $M$, and the incoming and outgoing lepton four-momenta are $k_\mu = (E, \vec{k})$ and $k'_\mu = (E', \vec{k}')$, respectively.

This expression assumes the exchange of a single virtual photon between the incoming lepton and the hadron. The differential cross-section is defined in terms of the so-called leptonic tensor $l_{\mu\nu}$ and the hadronic tensor $W^{\mu\nu}$. The virtual photon probing the hadron structure carries four-momentum $q_\mu=k_\mu-k'_\mu$. The Bjorken variable is defined as
\begin{equation}
x=\frac{Q^2}{2P\cdot q} \, ,
\end{equation}
where $0 \leq x \leq 1$ corresponds to its physical range. In the DIS limit $Q^2$ becomes very large, while $x$ is kept fixed. For a spin-$1/2$ baryon one may write the following decomposition for the hadronic tensor \cite{Anselmino:1994gn,Lampe:1998eu}
\be
W_{\mu\nu}=W^{\text{(S)}}_{\mu\nu}(q,P)+i \,
W^{\text{(A)}}_{\mu\nu}(q,P,S) \, , 
\ee
where the (Lorentz-index) symmetric part $W^{\text{(S)}}_{\mu\nu}$ includes the spin-independent structure functions $F_1(x, Q^2)$ and $F_2(x, Q^2)$, and the spin-dependent ones $g_3(x, Q^2)$, $g_4(x, Q^2)$ and $g_5(x, Q^2)$. On the other hand, the (Lorentz-index) antisymmetric part $W^{\text{(A)}}_{\mu\nu}$ in the general expression contains the so-called anti-symmetric structure functions $g_1(x, Q^2)$, $g_2(x, Q^2)$ and $F_3(x, Q^2)$.

Using the optical theorem, which relates the forward Compton scattering amplitude to the DIS cross section, it follows  
\begin{equation}
W_{\mu\nn}^{({\text{S}})} = 2 \pi \
{\text{Im}}\left[T_{\mu\nu}^{({\text{S}})}\right]
\,\,\,\,\, {\text{and}} \,\,\,\,\,\   W_{\mu\nu}^{({\text{A}})} = 2 \pi \ {\text{Im}}\left[T_{\mu\nu}^{({\text{A}})}\right] \, ,
\end{equation}
with
\begin{equation}
T_{\mu\nu}\equiv i \int d^{4}x \ e^{i q\cdot x} \langle P|
{{\hat{\text{T}}}} \{J_\mu^{\text{em}}(x) J_\nu^{\text{em}}(0)\} |P \rangle \, ,
\end{equation}
where $J_\mu^{\text{em}}$ represents the electromagnetic current inside the hadron state $|P \rangle$.

In QCD the functions $g_3$, $g_4$, $g_5$ and $F_3$ do not appear for electromagnetic DIS. However, considering an IR deformation in ${\cal {N}}=4$ supersymmetric Yang-Mills theory, $F_3$ is non-zero \cite{Hatta:2009ra,Kovensky:2017oqs,Kovensky:2018xxa}. In this specific situation massless Nambu-Goldstone modes appear from the spontaneous breaking of the $R$-symmetry \cite{Hatta:2009ra} of ${\cal {N}}=4$ SYM. It allows for a contribution to the $g_1(x, Q^2)$ structure function which is obtained by using the relation $g_1(x, Q^2)=F_3(x, Q^2)/2$. For more details of these calculations we refer the reader to references \cite{Hatta:2009ra,Kovensky:2017oqs,Kovensky:2018xxa}.

QCD and ${\cal {N}}=4$ SYM are different theories, 
specifically ${\cal {N}}=4$ SYM theory contains non-Abelian $SU(N_c)$ gauge fields (which represent the gluonic sector of this theory), gaugino fields, and six real scalar fields, all transforming in the adjoint representation of the gauge group $SU(N_c)$. However, within the parametric regimes of $Q^2$ and $x$ that we are interested in, the dominant contribution for both theories to the DIS process comes from the gluonic sectors, which are similar in both theories. Therefore, the behavior of the BPST and the Holographic-$A$ Pomerons turns out to be universal, while the model dependence is related to the IR deformation and the hadron impact factor.

In the work \cite{Kovensky:2018xxa} it has been obtained the helicity structure function $g_1$. This equation was obtained assuming that the kernels for $j \approx 1$ (Reggeized gauge field exchange) and $j \approx 2$ (Reggeized graviton exchange) can be approximately described in the same way \cite{Kovensky:2018xxa}. There are important changes of this derivation with respect to the derivation of the symmetric function $F_2$, since in the $\tilde{t}$-channel there is a Reggeized gauge field  exchange instead of a Reggeized graviton. Therefore, for $t<0$ and $0<|t| \ll s$, which corresponds to the UV region of the gauge theory leads to $j_{{\text Max}}=1$. The corresponding expression for $g_1(x, Q^2)$ is
\begin{equation}
g^{{\text{A}}_4 {\text{Pomeron}}_{\text{HW}}}_1(x, Q^2)  = \frac{C \rho^{-1/2} \ e^{(1-\frac{\rho}{4}) \tau_b}}{\tau_b^{1/2}} \left(e^{-\frac{\log^2{(Q/Q')}}{\rho \tau_b}}+ {\cal{F}}(x, Q, Q') \ e^{-\frac{\log^2{(Q Q' z^2_0)}}{\rho \tau_b}}\right)  \, .
    \label{}
\end{equation}
Notice that the parameters $\rho$, $Q'$ and $z_0$ should be fixed by the fitting of $F^{\text{BPST}_{\text{HW}}}_2(x, Q^2)$ to data, since the physical meaning of them is the same in both structure functions. Then, there is only one free parameter to fit to $g_1^p$ data, the overall constant $C$. Details of this derivation are given in reference \cite{Kovensky:2018xxa}.

\end{appendices}

\pagebreak

\end{document}